# Effect of strain on the electronic structure of $La_{0.7}Ca_{0.3}MnO_3$

P.R. Sagdeo*, R.J. Choudhary and D.M. Phase

UGC-DAE Consortium for Scientific Research, University Campus, Khandwa Road, Indore-452001, INDIA.

**Abstract:** The effect of substrate strain on the electronic valence band structure of $La_{0.7}Ca_{0.3}MnO_3$ thin films has been investigated. For this purpose $La_{0.7}Ca_{0.3}MnO_3$ thin films have been *simultaneously* grown on $SrTiO_3$ and $LaAlO_3$ substrates using pulsed laser deposition technique. These samples were characterized by x-ray diffraction, x-ray reflectivity, core level x-ray photoelectron spectroscopy, and scanning electron microscopy measurements. Our experiments confirmed that all these samples have *same chemical composition* but *different strain* configuration. The electronic structure of these samples is probed through valence band spectroscopy measurements on Indus-1 synchrotron x-ray source. We observe that strain has large effect on the valence band electronic structure of $La_{0.7}Ca_{0.3}MnO_3$. The results are explained on the basis of strain induced orbital occupancy and change in the crystal field splitting due to Mn-O bond length.

* sagdeo@csr.ernet.in , prsagdeo@yahoo.com

**Introduction:**

The highly correlated electron systems such as $ABO_3$ type perovskite manganites are known for their *rich* electronic, magnetic and structural phase diagram and complex phase coexistence of these phases [1]. In these systems it appears that the spin and the orbital degrees of freedom associated with $e_g$ electron are responsible for the complex phenomenon [2]. In the recent past various structural, electronic and magnetic phase coexistences of varying length scales have been reported in these materials [3-10]. *The phase coexistence in manganites not only provides abundant physics but also is responsible for the observed colossal magnetoresistance in these materials. Therefore, from the technological point of view it is important to know the origin of phase coexistence in manganites, in order to tailor the electronic properties of these materials.* The theoretical studies answering the issue of phase coexistences of various length scales have been reported in literatures [11-17]. The origin of nano-meter scale phase-coexistence has been attributed to the chemical inhomogeneity at A and B sites (A type rich or B type rich regions), the random-field Ising models, and also by considering two types of electronic wave functions (L-B model). However, theory incorporating primarily the structural aspects [18]; like elastic energy landscapes (strain-fields), predicts both nano-meter and micrometer sized phase coexistence. Thus, it is argued that the strain plays very crucial role in these manganites and is responsible for various electronic, structural and magnetic phases observed in these materials [18]. Therefore, if in a manganite system there is coexistence of different strain fields, the system would have coexistence of different electronic structure. To the best of our knowledge there is no clear experimental evidence, which demonstrates the effect of strain on electronic structure of these manganites. Therefore, keeping in view the crucial role of strain on the electronic structure, we aim to probe the electronic structure of these manganites having different strain configuration. These different strain configurations are achieved by *simultaneously* depositing the thin films of $La_{0.7}Ca_{0.3}MnO_3$ on different substrates such as, $LaAlO_3$ (LAO) and $SrTiO_3$ (STO) so that the films would experience compressive strain and tensile strain on these chosen substrates respectively.

**Experimental:**

The polycrystalline bulk target of $La_{0.7}Ca_{0.3}MnO_3$ (LCMO) has been prepared by standard solid-state reaction route [19]. This target is used to grow thin films of LCMO on c-axis oriented LAO and STO substrates. In order to get same chemical composition on each substrate, thin films were grown *simultaneously* on ~5mm x 5mm substrates. During the deposition, the oxygen partial pressure was maintained at 350mT, the target to substrate distance was kept at 4.5 cm and the substrate temperature was maintained at $650^0C$. These as grown samples were well characterized by x-ray diffraction (XRD), x-ray reflectivity (XRR), scanning electron microscopy (SEM) and core level x-ray photoelectron spectroscopy (XPS). The electronic valence band structure of these samples is probed through valence band spectroscopy (VBS) measurements carried out using BL-2 beam line on Indus-1 synchrotron radiation source [20]. Valence band spectra were collected using the Omicron energy analyzer (EA 125).

**Results and discussion:**

The XRD studies suggest the c-axis oriented epitaxial growth of LCMO on these substrates. The SEM studies suggest the uniform surface morphology of LCMO on these substrates. The thicknesses of the grown films were determined using XRR measurements and it was ~ 300Å for both samples. In Figure 1 we show the core level x-ray photoelectron spectroscopy spectra collected using Aluminum Kα radiation (1486.6eV) of both the samples. It is clear that both the spectra overlap. In order to estimate and compare the chemical composition for both the films we have carefully calculated the area under each peak taking in to account the atomic sensitivity. We found the same values (within error bar) for area under the peaks for both the samples. This confirms that all the grown samples have same chemical composition.

The valence band structures of these well-characterized samples were studied using measurements in the energy range of 40eV to 70eV. Figure 2 (a) shows the representative valence band spectrum (VBS) for sample deposited on STO substrate collected using incident photon energy of 52eV. This is resonant energy for Mn 3p-3d transition and all the features of VBS are clearly visible at this energy [21]. For all the measurements, the sample position (x, y, z) and the angle of incident beam (φ: angle between the surface of the sample and incident beam) were kept constant. From the figure-2 (a) it is clear that the observed VBS can be divided in to five peaks, as labeled from A to E. The intensity features at A and B are due to Mn-3d character [21]. In perovskite manganites the Mn ion is surrounded by six oxygen anions ($O^{-2}$) in octahedral cage, this gives rise to the splitting of degenerate 3d orbital of $Mn^{+3}$ in to $e_g(\sigma)$ and $t_{2g}(\Pi)$ levels. The peak A and B are attributed to the $e_g(\sigma)$ and $t_{2g}(\Pi)$ levels [21]. The peaks C and D are due to the O-2p character [21,22]; the feature C is assigned to the O-2p ($\Pi$) and D to O-2p ($\sigma$) character respectively. The detailed valence band study on these samples is reported by Zhang et al. [21].

Figure 2 (b) shows the valence band data (normalized with post mirror current) for all the studied samples. From the figure it is clear that there is a large difference between the VBS of these samples. It is important to note that all the samples were grown *simultaneously,* and have same chemical composition (figure-1); therefore, it appears that the observed difference in the valence band spectra must be due to the different strain provided by substrates. It is important to note that *Ma et al* [23] have theoretically predicted correlation effect of strain on the electronic structure of $La_{0.67}Sr_{0.33}MnO_3$ through *ab-initio* studies and obtained the similar results. The obtained VBS are de-convoluted, and the difference $\Delta(t_{2g}-e_g)$ is obtained for both the samples. We observed that the numerical values of these parameters vary from sample to sample. It appears that this difference may be due to the different strain configurations of these samples.

The lattice parameters for powder sample of $La_{0.7}Ca_{0.3}MnO_3$ (*Pbnm* space group) are a = 5.451, b = 5.4668 and c = 7.70 Å, where as the lattice parameters of $SrTiO_3$ and $LaAlO_3$ are 3.90 and 3.78 Å respectively. Therefore, it is natural to expect that the c-axis oriented LCMO will grow on the a-b plane along the diagonal of substrate unit cell and will occupy four such unit cells. Thus the film grown on STO and LAO substrates will have in plane expansive strain and compressive strain respectively. In the case of perovskite manganites six oxygen atoms surround the Mn atom in an octahedral cage, thus film on the STO substrate will have longer in plane Mn-O bond length than that of the film on LAO substrate. Thus the potential experienced by Mn ion due to neighboring four in plane oxygen anions will be more in the case of LAO substrate than that of the STO substrate. Therefore, there will be more crystal field splitting of Mn-3d i.e. $\Delta(t_{2g}-e_g)$ in LCMO on LAO substrate than that on the STO substrate. Table 1 summarizes the crystal field parameters obtain from the VBS. It appears that the binding energy associated with $e_g$ (itinerant) electron and the difference $\Delta(t_{2g}-e_g)$ critically depend on the strain condition. The present experimental studies reveal that the strain has large effect on the valence band structure. Therefore, the co-occurrence of different types of strain in a system will lead to the electronic phase separation as suggested by Ahn et al [18].

In summary the effect of substrate strain on the valence band electronic structure of $La_{0.7}Ca_{0.3}MnO_3$ has been investigated, through photoelectron spectroscopy. Our experimental findings suggest that the strain has large effect on the valence band electronic structure of these manganites. The substrate primarily modifies the extent of crystal field splitting and also has large effect on the binding energy of $e_g$ charge carriers. The study reveals that it is possible to tune the electronic properties of manganites by controlling the strain parameter.

**Acknowledgement:**

Authors sincerely thank Mr. A. Wadikar for valuable help during experiment. Authors are also thankful to Dr. P. Chaddah and Prof. Ajay Gupta for their encouragement. One of the authors PRS thanks DST Government of India for financial support.

**Figure captions:**

**Figure 1:** Core level XPS spectra for $La_{0.7}Ca_{0.3}MnO_3$ deposited on LAO (red line) and STO (black line) substrates. The inset shows the slow scan for individual elements La, Ca, Mn and Oxygen on STO and LAO substrates.

**Figure 2(a)**: Representative Valence band spectra for $La_{0.7}Ca_{0.3}MnO_3$. The spectra can be divided in to five peaks, as labeled from A to E.

**Figure 2 (b)**: Normalized valence band spectra for $La_{0.7}Ca_{0.3}MnO_3$ deposited on LAO (red line) and STO (black line) substrates; difference in the valence band spectra is clearly visible. The inset shows the variation of density of states near the Fermi-level $E_f$.

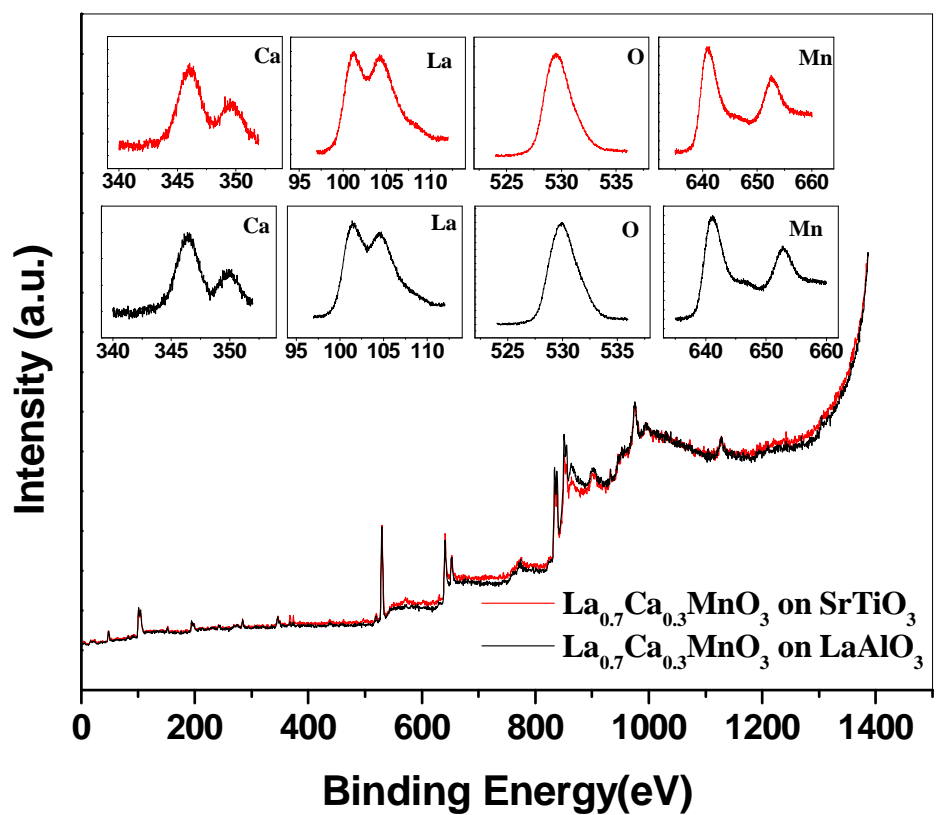

**Figure 1**

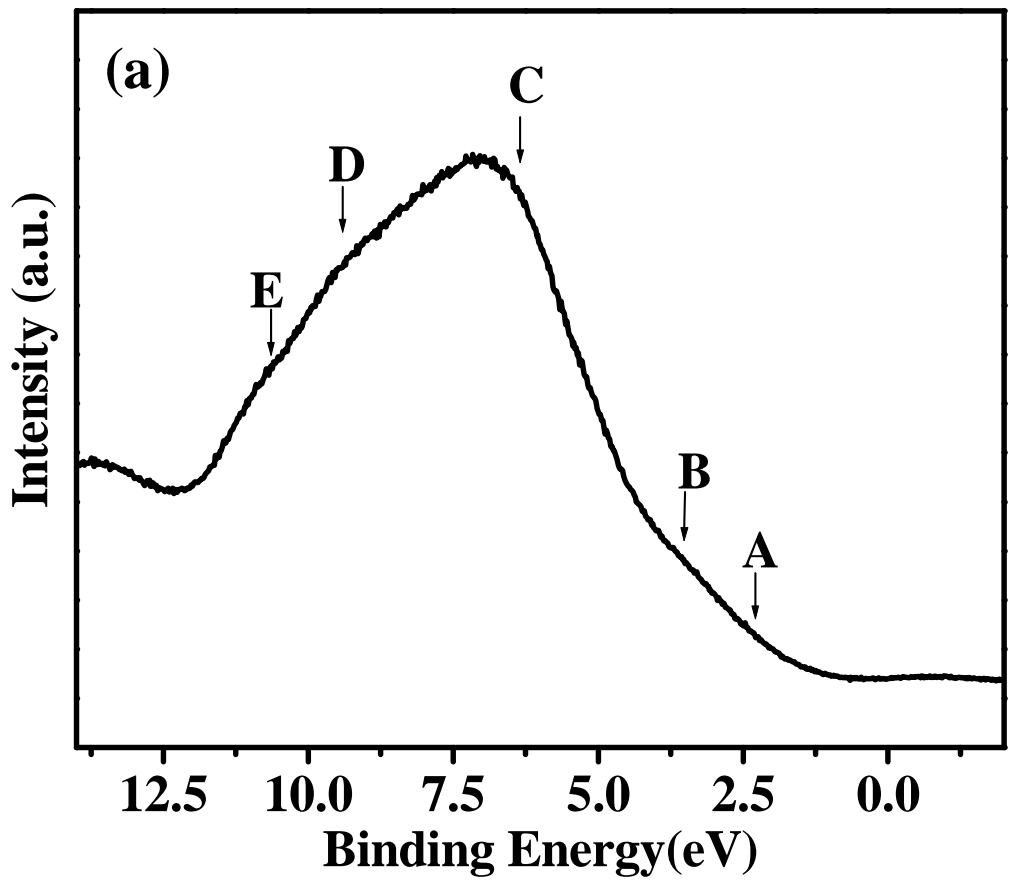

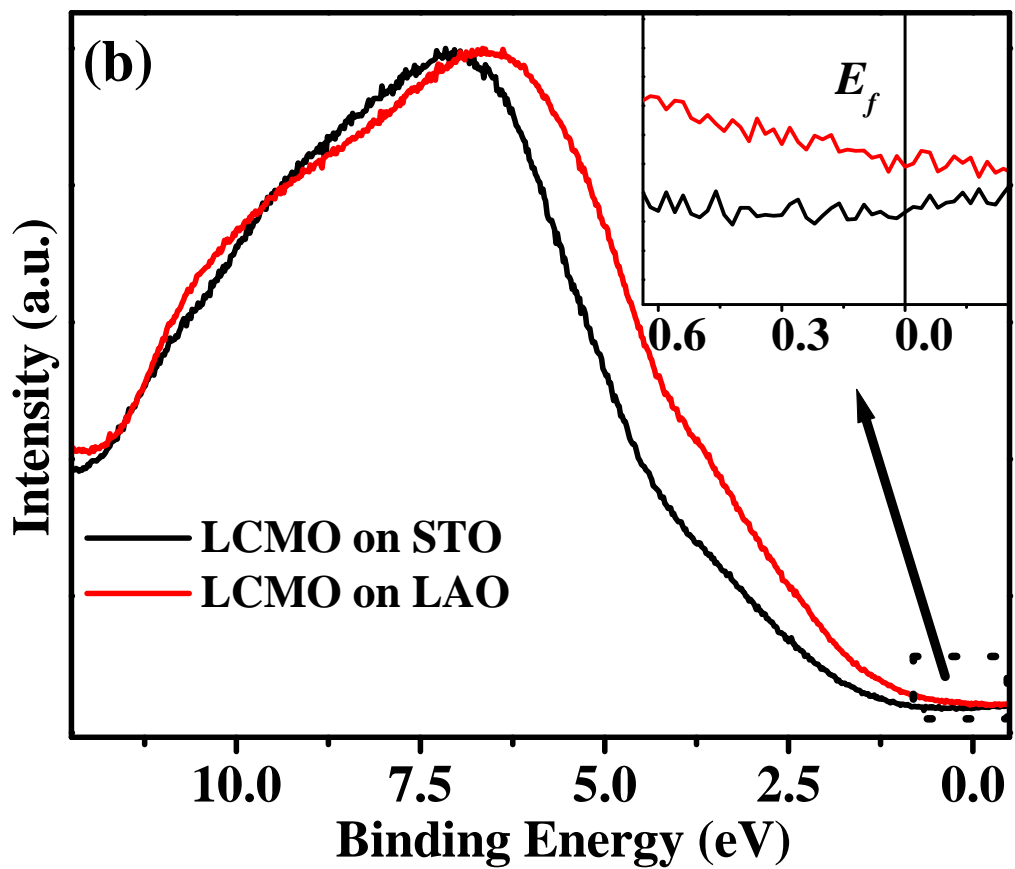

Figure 2

**Table1:** Crystal field parameters obtain from valence band spectra and the area under the curves obtain from core level spectroscopy for LCMO samples deposited on LAO and STO substrates.

| Sample | $t_{2g}$ | $e_g$ | $\Delta (t_{2g}-e_g)$ | Area Mn | Area La | Area Ca | Area O |
|---|---|---|---|---|---|---|---|
| $La_{0.7}Ca_{0.3}MnO_3$ on LAO | 3.60 | 2.10 | 1.50 | 16.90±0.1 | 9.27±0.12 | 9.43±0.12 | 5.55±0.1 |
| $La_{0.7}Ca_{0.3}MnO_3$ on STO | 3.45 | 2.30 | 1.15 | 16.86±0.1 | 9.33±0.12 | 9.39±0.12 | 5.51±0.13 |